\documentclass[reprint,
 twocolumn,
 amsmath,amssymb,
 aps,
 bbm,
groupedaddress
]{revtex4}

\usepackage{graphicx}
\usepackage{dcolumn}
\usepackage{bm}
\usepackage{bbm}

\usepackage{graphicx}
\usepackage{comment} 
\usepackage{color}
\usepackage{mathtools}
\usepackage{amsmath}
\usepackage{amssymb}
\usepackage{amsfonts}
\usepackage{empheq}
\usepackage{bbm}
\usepackage{MnSymbol}
\usepackage[dvipsnames]{xcolor}
\usepackage[normalem]{ulem}

\usepackage[colorlinks]{hyperref}
\AtBeginDocument{%
  \hypersetup{
    citecolor=Magenta,
    linkcolor=Red,   
    urlcolor=Black}}

\begin{document}

\preprint{APS/123-QED}

\title{Statistical Mechanics of Semantic Compression}

\author{Tankut Can}
\email{tankut.can@gmail.com}

\affiliation{Department of Physics, Emory University, Atlanta, GA}%

\date{\today}
\begin{abstract}

The basic problem of semantic compression is to minimize the length of a message while preserving its meaning. This differs from classical notions of compression in that the distortion is not measured directly at the level of bits, but rather in an abstract semantic space. In order to make this precise, we take inspiration from cognitive neuroscience and machine learning and model semantic space as a continuous Euclidean vector space. In such a space, stimuli like speech, images, or even ideas, are mapped to high-dimensional real vectors, and the location of these embeddings determines their meaning relative to other embeddings. This suggests that a natural metric for semantic similarity is just the Euclidean distance, which is what we use in this work. We map the optimization problem of determining the minimal-length, meaning-preserving message to a spin glass Hamiltonian and solve the resulting statistical mechanics problem using replica theory. We map out the replica symmetric phase diagram, identifying distinct phases of semantic compression: a first-order transition occurs between phases marked by the emergence of paraphrases, whereas a continuous crossover is seen from extractive to abstractive compression. We speculate on which features of the phase diagram are captured by replica symmetry, and which features change under replica symmetry breaking. We conclude by showing numerical simulations of compressions obtained by simulated annealing and greedy algorithms, and argue that while the problem of finding a meaning-preserving compression is computationally hard in the worst case, there exist efficient algorithms which achieve near optimal performance in the typical case.

\end{abstract}

\maketitle

\section{Introduction}

Human working memory has a limited capacity, as revealed from numerous experiments using unstructured stimuli \cite{cowan2012working}. Nevertheless, we have the ability to process information on extremely long timescales, in apparent contradiction to this finite capacity. The crucial ingredient for accomplishing this is compression, often called ``chunking" in cognitive science, whereby stimuli are recoded into more compact representations \cite{miller1956magical}.

Compression is routinely observed in social communication. Bartlett \cite{Bartlett1932} showed that when stories are transmitted between humans, they tend to become shorter and more stereotyped. Furthermore, in experiments on human memory for narratives, in which subjects read a story and are subsequently asked to retell it, there is a strong tendency to produce a compressed version of the story in the retelling, using efficient paraphrases and summaries \cite{Musz2022, Georgiou2023,zhong2024random}.

Importantly, human communication involves {\it lossy} compression. For instance, in the narrative memory experiments described above, the verbatim text of the original story, also known as the surface structure, cannot be reconstructed from the recall of participants. If the surface structure does not seem to matter, then what is being transmitted during communication? A natural hypothesis is that {\it meaning} is the important thing, and surface structure can be sacrificed as long as meaning is kept invariant. 

But what is meaning? This notoriously elusive concept finds its most concrete formulation in the study of semantics, which seeks to understand how meaning arises in language \cite{Jackendoff2002}. Far from being a mere formal exercise, the study of semantics is central to the psychology of human memory. Short-term sentence recognition experiments show that details of the wording of a sentence are easily forgotten, while the meaning or ``gist" of a sentence is kept much longer and more stably in memory \cite{Gomulicki1956, Sachs1967,fillenbaum1966memory}. This can be seen by testing paraphrases which preserve meaning, against sentence variants which change the original meaning. Measuring forgetting over longer timescales (days to months) confirms this observation, showing that memory for surface structure decays much faster than memory for semantic structure, which includes higher-level abstractions of a text that together give it meaning to an individual \cite{kintsch1990sentence,kintsch1998comprehension}. In short, our memory for discourse appears to be primarily semantic in nature. Therefore, while the compression involved in human communication may lose surface structure, it tends to preserve semantic structure; for this reason, we refer to this process as {\it semantic compression}.

Semantic compression therefore plays a central role in human communication. Traditionally, lossy compression is the purview of rate-distortion theory, and semantic compression has been studied precisely in this context \cite{Guler2014,Liu2021}. Furthermore, there have been more general theories which frame pragmatic communication between agents in terms of optimal transport \cite{wang2020mathematical}. While necessary, these general theoretical frameworks leave open some questions about the mechanism of compression in particular settings. For instance, what is the interplay between the structure of representations in semantic space, and the capacity for compression? To address such a question, we must try to make explicit contact with the representations that humans (and machines) make use of. In other words, we must specify the semantic distortion function.

To define the distortion function, we take inspiration from both machine learning and cognitive neuroscience. Lexical items, such as words and phrases, are clearly stored in long-term memory \cite{Jackendoff2002}. Studying brain responses during a story listening task, researchers were able to map out what they called a semantic network in the brain \cite{Tang2022, Kumar2021}. This was then used to argue that semantic representations are continuously represented in brain activity in the whole cortex \cite{Huth2012}. 

The idea of semantic spaces has a long history in psychology, dating back to the spreading activation theory in the late 60's \cite{quillian1967word}, and later \cite{collins1975spreading}. Similarly, G\"ardenfors \cite{Gardenfors2014} makes the argument for continuous conceptual spaces in neural population representations. Semantic spaces were also studied in \cite{vigano2021grid}, and argued to be closely related to the brain's native instruments for representation of spatial geometry. This mirrors the ability of hippocampus to generate cognitive maps which represent abstract categories instead of spatial location \cite{aronov2017mapping}. The question remains: how does the brain represent the presumably high-dimensional spaces involved in our semantic knowledge base? Indeed, it was been argued that low-dimensional spaces are insufficient to describe the geometry of concepts \cite{tversky1986nearest}, and that complex networks \cite{steyvers2005large} or high-dimensional distributed representations \cite{bhatia2019distributed} are needed to account for semantic similarity judgements. A recent review has even argued in favor of explicit vector space representations of concepts \cite{Piantadosi2024}.

In parallel research, language modeling in machine learning has naturally come upon the idea of using continuous vector spaces to represent word meanings. Some particularly vivid examples of this come from algorithms such as word2vec \cite{Mikolov2017} or GloVe \cite{Pennington2014}, which map each word or token in a lexicon to an {\it embedding vector} which lives in a high-dimensional Euclidean space. Remarkably, meaningful relations between words, such as analogies, are found to be encoded in geometric relations between their vector embeddings. For instance, the vector sum of 'royal' and 'man' is close to 'king' and 'prince' (albeit also close to other seeming non-sequiturs). Similarly, analogies can be represented by vector addition, as with, ${\bf v}({\rm duke}) - {\bf v}({\rm male}) \approx {\bf v}({\rm duchess})- {\bf v}({\rm female})$. More generally, contrastive representation learning is an approach in deep learning that seeks to capture semantic similarity between any inputs (words, sentences, images) in an embedding space using continuously differentiable embeddings \cite{Le-Khac2020}. In a real sense, deep learning rests on the power and efficacy of a continuous semantic space. 

Given the stunning success of language modeling, and the evidence from cognitive neuroscience, we assume that the space of meaning, or semantic space, is given by a Euclidean vector space. Furthermore, we will assume that two meanings are similar if they are close in Euclidean distance in this semantic space. Every message, no matter its length, will be represented in this semantic space by a vector. We further assume a ``bag of words" representation for every message, in which the embedding of a long message is just the linear sum of the embeddings of all of its constituent lexical items. This allows us to define the problem of semantic compression as one which minimizes the Euclidean distance between the embeddings of two messages, subject to the constraint that the compressed message has fewer constituent tokens than the target message. All of these objects will be defined mathematically below. 

In this paper, we introduce a statistical mechanical model for semantic compression, and present its mean-field phase diagram under the replica symmetric (RS) Ansatz. We show how the RS theory reveals that qualitatively different phases of compression are encountered as one varies embedding dimension $P$, lexicon size $N$, target message length $L$, and compressed length $\bar{L}$. In particular, we find a first-order transition in the RS theory between a phase with {\bf unique} compression to a phase with multiple compressions, or {\bf paraphrases}. We also find that the minimum distortion drops to zero in the paraphrastic phase. In the lossy phase with a unique compression, we can identify regimes, related by a crossover behavior, in which the compression is either: 1) {\bf extractive}, wherein the target can only be compressed by removing words or tokens; 2) {\bf abstractive}, in which multiple words in the target message are represented by a single word that does not appear in the target. We compare our phase diagram and order parameters to numerical experiments, showing where agreement is good, and in which regimes agreement breaks down. RS theory describes extractive lossy compression very well, but fails at capturing the details of the transition to lossless paraphrastic compression. Furthermore, we find a strictly positive lower bound on the minimal distortion using the annealed free energy, suggesting a breakdown of the qualitative features implied by the RS ansatz across this transition. Finally, we compare a costly Monte Carlo minimization of the distortion, to an efficient greedy algorithm that minimizes the distortion one token at a time by always finding the next closest token embedding. Remarkably, for the typical case scenario we study in this paper, the greedy algorithm is nearly optimal, and finds solutions that are well described by RS MFT in certain regions of the phase diagram.

\section{Mathematical Model of Semantic Compression}

Here we lay out a set of simplifying assumptions that will allow us to introduce a tractable model of semantic compression. 

{\bf Assumption 1:}  The {\it semantic space} is Euclidean space $\mathbbm{R}^{P}$. 

Humans have at their disposal a large lexicon of words, word fragments, and phrases, all of which are stored in long-term memory and used in a combinatorial manner (e.g. via grammar) to construct messages. We denote the lexicon with $N$ items by $\mathcal{L}_{N}$, and denote the lexical items or listemes \cite{Jackendoff2002} by $s_{i} \in \mathcal{L}, \quad i = 1, ..., N$. A message of length $L$ is defined as a sequence of $L$ lexical items 
\begin{align}
S({\bf k}) = s_{k_{1}} s_{k_{2}}... s_{k_{L}}, \quad |S| = L. 
\end{align}
From here we see that every message can be represented by a vector ${\bf k} \in \mathbbm{Z}_{N}^{L}$, where $k_{1}$ is the integer label of the first listeme in the message, and so forth. It is useful to represent a message by a {\it count vector} $c(S) \in \mathbbm{Z}^{+ N}$, where each entry $c_{i}$ is a positive integer that gives a count of the number of times listeme $s_{i}$ appears in $S$:
\begin{align}
c_{i} = \sum_{j = 1}^{L} \delta_{i k_{j}} . \label{eq:count_vec_components}
\end{align}
Henceforth, we drop the argument of $c = c(S)$ and simply refer to vector $c$ as the message. 

Every message has a representation in semantic space $\mathbbm{R}^{P}$. If we denote the space of all messages $\mathcal{S}$, then we define an {\it semantic embedding} function which maps every message to a point in semantic space:
\begin{align}
X: \mathcal{S} \to \mathbbm{R}^{P}.
\end{align}
We assume that each individual lexical item has a unique semantic embedding $X(s_{i}) = E_{i}$. To make progress with our model, we make the next crucial simplifying assumption

{\bf Assumption 2:} The semantic embedding of a message is a linear sum of the embeddings of each constituent lexical item, i.e.
\begin{align}
X(S({\bf k})) = \sum_{j} X(s_{k_{j}}) = \sum_{j = 1}^{L} E_{k_{j}} = \sum_{i = 1}^{N} c_{i} E_{i}. 
\end{align}
After the last equality, we have given a representation in terms of the count vector. We primarily use this representation in the rest of the paper. 

This assumption states that the meaning of an item in a message is independent of the structure of the message, i.e. independent of context. In natural language processing, this representation is usually referred to as a ``bag-of-words", since the embedding only depends on the set of words or listemes used, and is insensitive to the order or general context in which they are used. For example, with a bag-of-words representation, we cannot semantically distinguish ``Dog bites man" from ``Man bites dog", since these are composed of the same lexical elements. It is of course possible to expand the lexicon to include compound phrases like ``(dog, subject)" as well as ``(dog, agent)", but this would potentially lead to an unbounded growth of the lexicon, due to combinatorial explosion. Indeed, this endeavor amounts to capturing the infinite generative power of syntax with a fixed lexicon, which seems both unfeasible and extremely inelegant. Therefore, we fully acknowledge that our embedding has some obvious shortcomings, but nevertheless pursue the consequences to the end. We will find that even with this simple choice, there is a rich structure to semantic space. Inclusion of syntax and context more generally must be reserved for future work. 

The next assumption we make concerns the metric of semantic similarity. 

{\bf Assumption 3:} We take semantic dissimilarity, or equivalently semantic distortion, between two messages to be quantified by the Euclidean distance squared between their embeddings, 
\begin{align}
d(S, S') = || X(S) - X(S')||^{2}. \label{eq:distortion}
\end{align} 
 
A small distortion arises when semantic embeddings are close in semantic space. The most straightforward generalization of this entails imbuing the semantic space with a nontrivial metric. For instance, it has been argued that olfaction utilizes a hyperbolic embedding space \cite{Zhou2018}. In many pre-trained machine learning embeddings (e.g. OpenAI, SBert, etc.), the semantic space is the unit $P-$sphere, in which case the distortion function is equivalent to the cosine similarity ${\rm cos}(S, S')$ after a shift, i.e. $d(S, S') = 2 - 2{\rm cos}(S, S')$.

The preceding assumptions concern the structure of semantic space and semantic similarity. The next few assumptions concern the structure of embeddings and the messages.

{\bf Assumption 4:} The vector embeddings $E_{i}$ are random Gaussian vectors with the following moments
\begin{align}
{\rm E} \left[ E_{i}^{\mu} \right] = b^{\mu}, \quad {\rm E}\left[ (E_{i}^{\mu} - b^{\mu}) (E_{j}^{\nu} - b^{\nu}) \right] = \delta_{\mu \nu} \Sigma_{ij}. \label{eq:embeddings}
\end{align}

The finite mean value affects all embeddings in essentially the same way. However, this uniform shift will have consequences on the overall structure of the embeddings, and consequently on the likelihood of finding efficient compressions or paraphrases. The nontrivial variance is supposed to reflect the fact that embeddings of semantically similar lexical items will tend to be correlated. Thus, under an appropriate indexing of the lexicon, $\Sigma_{ij}$ is expected to have a block diagonal structure. In the main discussion below, we specialize to the case of uncorrelated embeddings $\Sigma_{ij} = \delta_{ij}$, with zero mean $b^{\mu} = 0$. We explore the general setting in an upcoming paper. 

The assumption of random embeddings is not unreasonable. In fact, it can be observed that word embeddings from popular algorithms (e.g. word2vec or GloVe) appear to have components which follow a Gaussian distribution. These algorithms assume randomly initialized vectors assigned to each word, which are subsequently updated by some learning rule. It it conceivable that this learning rule does not change the distribution, but the relative position of the random vectors. This means that any collection of random vectors can be used as embedding vectors - the only question is which word gets assigned to which vector. This fascinating argument was made in \cite{nurmukhamedov2022hyperbolic}, but for random points in hyperbolic space instead of high-dimensional Euclidean space. 

Finally, we constrain the space of messages:

{\bf Assumption 5:} The components if the count vector defined in Eq. \ref{eq:count_vec_components} are binary, i.e.  $c_{i} \in \{0,1\}$.

This means messages are not permitted to have any repeated lexical items $s_{i}$. If we treat the lexical items as words in an actual text, then this assumption is obviously wrong. However, we may treat the lexicon not as representing individual words but unique concepts. In this case, it is a little more sensible to have a message that does not have repeating concepts.

Having laid out the essential details of the semantic space and the embedding function, we are now in a position to define a statistical mechanics of semantic compression.

\section{Statistical Mechanics of Semantic Compression}

We will ultimately be interested in the thermodynamic scaling limit, which involves taking the lexicon size to infinity. Since for our bag-of-words embedding function, each message is represented uniquely by the vector of counts $c$, we henceforth denote the original message by $c$, and the compressed message by $\bar{c}$. The 1-norm of the counts vector gives the total length of the message, which we denote as $|| c||_{1} = L$ and $|| \bar{c}||_{1} = \bar{L}$ for the original and compressed message, respectively. The Hamiltonian is 
\begin{align}
H(c, \bar{c}) = \frac{1}{2 N} d(c, \bar{c}) = \frac{1}{2N} \sum_{i, j} \sigma_{i} J_{ij} \sigma_{j}, \label{eq:H}
\end{align}
where $\sigma_{i} = c_{i} - \bar{c}_{i}$, and $J_{ij} = E_{i} \cdot E_{j} \equiv \sum_{\mu = 1}^{P} E_{i}^{\mu} E_{j}^{\mu}$. With this Hamiltonian we can proceed to define a statistical mechanics formulation of our combinatorial optimization problem \cite{Parisi1987}. For quenched embeddings $E_{i}$ and original message $c$, we must find the optimal compression $\bar{c}$ of a fixed length $\bar{L}$. Therefore, we define the partition function
\begin{align}
Z( E, c, \bar{L}) = \sum_{ \bar{c}_{i}, || \bar{c}||_{1} = \bar{L}} \exp \left( - \beta H(c, \bar{c}) \right), \label{eq:Z}
\end{align}
where the sum is constrained to be over all messages $\bar{c}$ of a fixed length $\bar{L}$. From this, we obtain the free energy density
\begin{align}
f_{\beta}(L, \bar{L}, b,\Sigma) =  - \frac{1}{\beta N} \mathbb{ E} \left[  \log Z( E, c, \bar{L})  \right]_{E, c, || c||_{1} = L}.
\end{align}
Here again we take a constrained average over original messages $c$ at a fixed length $||c||_{1} = L$. We also average over random embeddings. 

The interesting question that our model is supposed to address is how the structure of embeddings is implicated in the ability to produce efficient semantic compressions. For trivial correlations in Eq. \ref{eq:embeddings}, $\Sigma_{ij} \propto \delta_{ij}$, the only relevant structure is the dimension of the embedding space $P$, and the size of the lexicon $N$. Therefore, we consider how the compression scales with their ratio
\begin{align}
\alpha = \frac{P}{N}, \label{eq:alpha_def}
\end{align}
which we refer to as the {\it relative embedding dimension}. Of course, the size of the original message and the target compressed length will interact with $\alpha$ to influence compressibility. Therefore, we define here the message length ratios.   
\begin{align}
\ell = \frac{L}{N}, \quad \bar{\ell} = \frac{\bar{L}}{N}.
\end{align}
We also introduce the compression ratio
\begin{align}
C = \bar{L}/L ,
\end{align}
as an important control parameter in our model. The thermodynamic limit in our model amounts to taking $P, N, L, \bar{L} \to \infty$ while keeping fixed $\alpha$, $\ell$, and $\bar{\ell}$. 

The average distortion in our model is just the mean energy density, and is given by
\begin{align}
D(\beta) = \partial_{\beta} \left( \beta f_{\beta}\right).
\end{align}
We can also find the minimum distortion from the zero temperature limit 
\begin{align}
D_{min} =\frac{1}{N} \mathbbm{ E}\left[ {\rm min}_{\bar{c}} H(c, \bar{c}) \right]_{E, c} =   \lim_{T \to 0} f_{\beta} .
\end{align}

\section{Order Parameters from Mean-Field Theory}

The calculation of the free energy is carried out by straightforward application of replica theory. We present the details in the supplemental material. Below, we give the main results and their interpretation. But first, we introduce the order parameters and provide an intuition for their meaning by studying a simple example. These order parameters are in fact quite natural in the context of spin glasses, but their meaning in the present context is not immediately apparent. 

The first order parameter is the average overlap between the original message and the compressed message:

\begin{align}
r= \mathbbm{ E}\left[ \frac{1}{N} \sum_{i = 1}^{N} c_{i} \langle \bar{c}_{i}\rangle \right]_{E, c}, \label{eq:overlap}
\end{align}
where we denote by brackets $\langle ... \rangle$ the average using the partition function (\ref{eq:Z}) (i.e. the thermal average), with $E_{i}$ and $c$ fixed (quenched). For binary counts $c_{i}, \bar{c}_{i} \in \{0, 1\}$, the overlap obeys the bounds
\begin{align}
 {\rm max}\left( 0, \ell + \bar{\ell} - 1\right) \le r \le  \bar{\ell}.
\end{align}
The upper bound is saturated when the compression is purely {\bf extractive}. This means that the only effective compression possible is one in which a subset of the original lexical items are used. An example of an extractive compression would be if ``The quick brown fox jumps over the lazy dog" was shortened to ``The fox jumped over the dog". Below this upper bound, the compression must utilize paraphrases, since it would require some of the words in $\bar{c}$ to not have appeared in $c$. A paraphrasing compression might look like ``A fast fox leaped over the canine". Approaching the lower bound requires an {\bf abstractive} compression, in which a majority of lexical items in the compression $\bar{c}$ are not in the original message $c$. For this well-known example sentence that uses every letter in the English alphabet, an efficient abstractive compression could be ``a famous pangram". 

There is also a shifted inter-replica overlap, similar to the Edwards-Anderson (EA) order parameter \cite{edwards1975theory}, characterizing the overlap between different ``ground-state" configurations:
\begin{align}
\Delta = \bar{\ell} -  \mathbbm{ E}\left[ \frac{1}{N} \sum_{i = 1}^{N} \langle \bar{c}_{i} \rangle  \langle \bar{c}_{i}\rangle \right]_{E, c}. \label{eq:EA}
\end{align}
Within the replica symmetric theory, $\Delta$ is single-valued and signals a phase transition. The range of the inter-replica order parameter for binary messages is given by
\begin{align}
0 \le \Delta \le {\rm min} \left( \bar{\ell}, 1 - \bar{\ell}\right).
\end{align}
The lower bound $\Delta = 0$ is saturated in the case that there is a {\bf unique compression} $\bar{c}$. When there are multiple states which achieve the minimal distortion, then $\Delta>0$. Since in this case, there are many ways to convey the same message without changing the meaning, we refer to this as the {\bf paraphrastic} phase. Note that positive $\Delta$ can occur both if there are exponentially many compressions, or just order one ground states whose overlaps are extensive in $N$. We cannot distinguish between these scenarios.  For the replica symmetric ansatz, we will see in what follows that for random Gaussian embeddings,  $\Delta = 0$ corresponds to the lossy compression phase, whereas $\Delta>0$ characterizes the lossless compression phase. However, we also establish that the minimal distortion is always positive throughout the bulk of the phase diagram (with the possible exception of some boundaries), so this lossy/lossless characterization is an artifact of the replica symmetric ansatz we use.

We can gain some intuition for these order parameters by considering first a simple limit of our model for semantic compression.

\subsection{Special Case: Weighted Hamming Compression}\label{sec:hamming}

For orthogonal patterns $E_{i} \cdot E_{j}  = w_{i} \delta_{ij}$ (which requires $P \ge N$), the Hamiltonian is the weighted Hamming distance between these bit strings
\begin{align}
H(c, \bar{c}) = \frac{1}{2N} \sum_{i = 1}^{N} w_{i} (c_{i} - \bar{c}_{i})^{2} .
\end{align}
The minimal distortion that is achievable is
\begin{align}
D_{min} = \frac{1}{N} \mathbbm{ E}\left[ \min_{\bar{c}}H(c, \bar{c}) \right]_{E, c} = \frac{\langle w \rangle }{2N} ( \ell - \bar{\ell}),\label{eq:min_hamming}
\end{align}
 which obtains when $\bar{c}_{i}$ is only nonzero for $i$ such that $c_{i} = 1$. In other words, $r = \bar{\ell}$ is saturated at its upper bound, and the compressions are strictly {\bf extractive}.

 Since each bit is unequally weighted, there will generically be a unique minimizer, which implies $\Delta= 0$. This situation is slightly different for the pure Hamming distance which has $w_{i} = 1$ for all $i$. In that case, there will be a large degeneracy of minimal distortion compressions, which leads to $\Delta >0$.
 
 We will see below that for random embeddings, the compression phase diagram is described by Hamming compression in the limit that $\alpha >>1$.

\section{Bounds on the Minimal Distortion}

We can use the annealed free energy to get a lower bound on the minimum distortion. Using Proposition 5.4 in \cite{mezard2009information}, we get

\begin{align}
\mathbb{E} \left[ \frac{1}{N} \min_{\bar{c}} H(c, \bar{c}) \right]_{ E,c} \ge  \max_{\beta} f_{\beta}^{a}
\end{align}
where the annealed free energy is

\begin{align}
f_{\beta}^{a} = - \frac{1}{\beta N} \log \mathbb{E} \left[ Z(E, c, \bar{L}) \right]_{E, c} . 
\end{align}

In the scaling limit of large $N$, we show in Appendix \ref{app:annealed} that the lower bound takes the form

\begin{align}
f^{a}_{\beta^{*}} =  - \frac{1}{\beta^{*}} \Phi(r^{*}, \beta^{*}), \label{eq:fa}
\end{align}
where 
\begin{align}
 \Phi(r, \beta) &= S(r) - \frac{\alpha}{2} \log \left( 1+   \beta ( \ell + \bar{\ell} - 2 r) \right), \\
 S(r) & = \ell h\left(r/\ell \right) + (1 - \ell) h \left( \frac{\bar{\ell} - r}{1 - \ell}\right), 
\end{align}
and $h(p) = - p \log p - (1 - p) \log (1 - p)$ is the binary entropy function. In Eq.\ref{eq:fa}, $r^{*}$ and $\beta^{*}$ are determined by both selecting the stable saddle-point:

\begin{align}
\Phi_{r}(r^{*}, \beta^{*}) &= 0, \quad \Phi_{rr}(r^{*}, \beta^{*}) < 0, 
\end{align}
and maximizing over $\beta$, which is equivalent to finding the zero annealed entropy solution:
\begin{align}
s^{a} = - \partial_{T} f^{a} & =  \Phi(r^{*}, \beta^{*}) -  \beta^{*} \Phi_{\beta}(r^{*}, \beta^{*}) = 0.
\end{align}

Evaluating these gives the bound in a compact and transparent form

\begin{align}
D_{min} \ge  \frac{ \alpha q^{*}}{2 (1 + \beta^{*} q^{*})} >0, \quad q^{*} = \ell + \bar{\ell} - 2 r^{*}.
\end{align}

Additionally, using $q^{*} \ge \ell - \bar{\ell} > 0$, we conclude that the annealed lower bound actually implies that the minimal distortion is finite for all compressed messages for $\alpha > 0$. Since the annealed entropy density diverges to negative infinite as $\beta \to \infty$, while it is finite and positive for $\beta = 0$, we know that $\beta^{*}$ should be finite positive number. Therefore, we cannot hope to completely eliminate distortion in the asymptotic limit. However, its value can come quite close to this annealed lower bound, as seen in Fig. \ref{fig:numerics}.

\section{Replica Symmetric Mean Field Theory}

\begin{figure*}[htbp!]

\includegraphics[width = .8\textwidth]{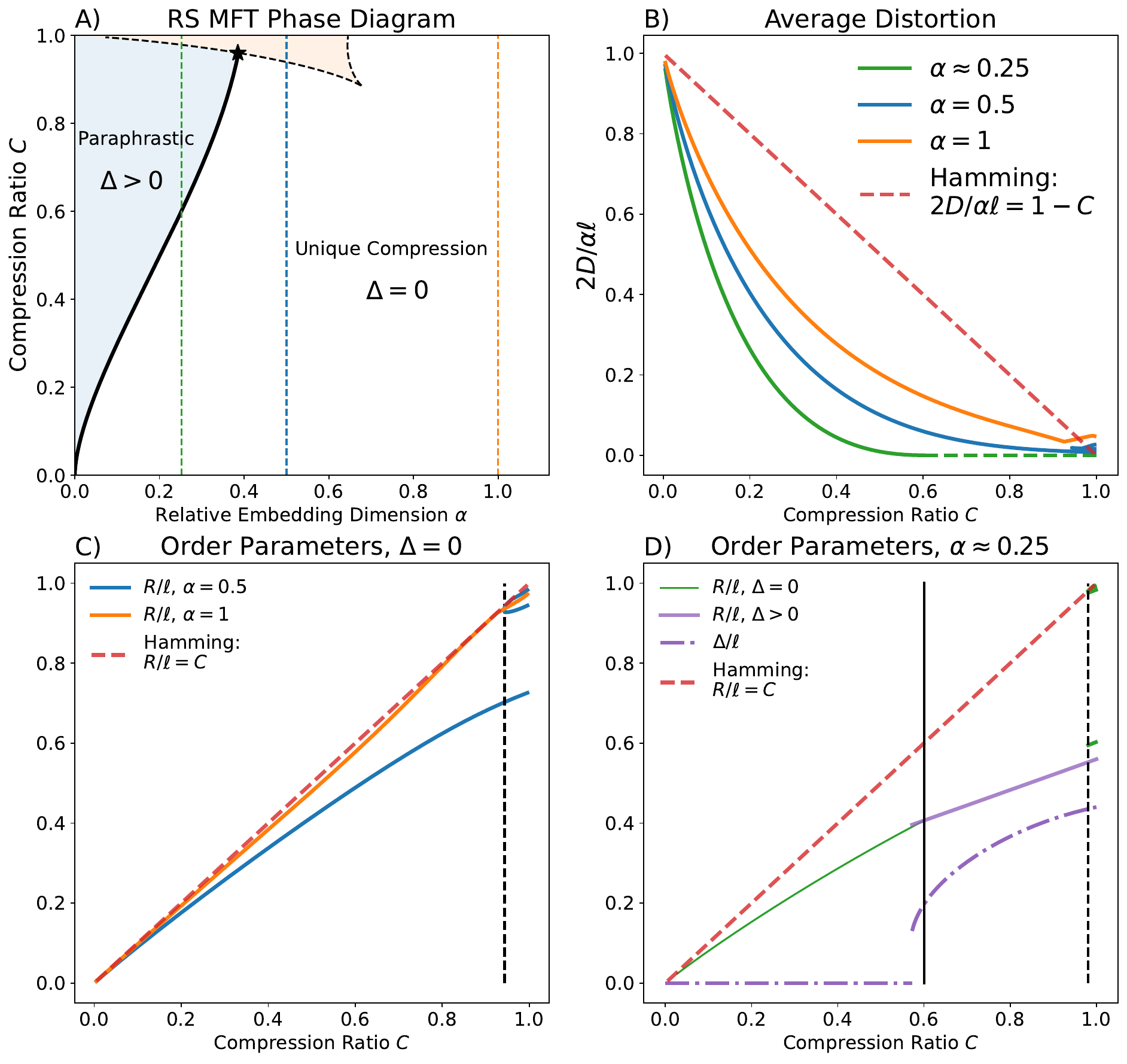}

	\caption{ {\it Semantic Compression Phase Diagram and Order Parameters} {\bf A) }The zero temperature phase diagram for the RS order parameters fixing $\ell = 0.4$. The discontinuous transition is indicated by the thick black line. In the paraphrastic phase (blue region), the inter-replica order parameter $\Delta  > 0$. Within the unique compression phase, $\Delta = 0$, there is a region (shaded orange, enclosed by black dashed curves) in which the RS MFT has multiple solutions. Outside of this region (white area), the RS MFT has a unique solution. Colored dashed lines show the slice along which the order parameters are computed in the other panels.
{\bf B)} Average distortion (at zero temperature) normalized by its value $\alpha \ell/2$ at $C = 0$. Outside the paraphrastic phase, the distortion never reaches zero, but generally decreases with compression ratio. For the green curve, the value of $\alpha \approx .251$ is chosen such that the RS distortion hits zero at exactly $C = 0.6$. {\bf C)} The overlap order parameters in the unique compression phase. For reference, we show the weighted Hamming compression limit (red dashed) in which $\Delta = 0$ and $r = \bar{\ell}$. For $\alpha = 0.5$ (solid blue), there is a bifurcation in the RS MFT at $C \approx 0.94$, above which two new solutions appear which are much closer to the Hamming limit. For larger $\alpha$, the overlap is close to the Hamming line for all compression ratios. {\bf D)} Overlap and inter-replica order parameter for $\alpha \approx 0.251$. There are no $\Delta = 0$ solutions to the order parameters in the region between the two vertical lines at $C = 0.6$ (solid black) and $C \approx 0.98$ (dotted black). However, $\Delta > 0$ solutions actually appear at compression ratios slightly smaller than $C = 0.6$. There is no reason to prefer one of these over the other in the RS theory, since they both have negative entropy.  Furthermore, the $\Delta>0$ solutions survive until $C = 1$. However, approaching $C = 1$ the RS MFT does yield a sensible physical solution, which we believe takes over for larger compression ratios. }
	\label{fig:phase}
\end{figure*}

The replica symmetric mean-field theory (RS MFT) is most naturally formulated in terms of the order parameter
\begin{align}
q_{ab} = \mathbbm{E} \left[ \frac{1}{N} \sum_{i = 1}^{N} \sigma_{i}^{a} \sigma_{i}^{b} \right]_{E, c, \bar{c}}.
\end{align}
This can be related in a straightforward way to the more ``physical" target-compression overlap (\ref{eq:overlap}) and inter-replica overlap (\ref{eq:EA}). In particular, for the replica symmetric ansatz, $q^{ab} = q_{0}\delta_{ab} + (q_{0} - q_{1})$, we have 
\begin{align}
q_{0} &= \ell + \bar{\ell} - 2 r, \quad q_{1} = q_{0} -\Delta . 
\end{align}
We review the full RS MFT in the supplementary material. For now, we present only the zero temperature ($\beta \to \infty$) limit. For small temperatures, $\Delta = \Delta_{0} + T \delta + O(T^{2})$. In the unique compression phase, $\Delta = 0$, $q_{0} \to q_{1} \equiv q$, and the RS MFT reduces to a set of two equations for $q$ and an auxiliary (Lagrange multiplier) variable $\lambda$:
\begin{align}
q &= \ell H_{1}(-\lambda) + (1 - \ell) H_{1}(\lambda), \label{eq:qbar_mft}\\
\ell - \bar{\ell} & = \ell H_{1}(-\lambda) - (1 - \ell) H_{1}(\lambda) , \label{eq:lambda_mft}
\end{align}
where
\begin{align}
H_{1}(\lambda) = \frac{1}{2}{\rm erfc}\left( \frac{\alpha/2 +\lambda}{\sqrt{2 \alpha q}}\right).
\end{align}
The second equation Eq. \ref{eq:lambda_mft} arises due to the hard constraint on the compression length $\bar{\ell}$. The mean distortion is equal to the minimal distortion in this limit, and given by
\begin{align}
D =  \frac{\alpha q}{2(1 + \delta)^{2}},
\end{align}
where
\begin{align}
\delta =\frac{ \ell H_{2}(-\lambda) + (1 - \ell) H_{2}(\lambda)}{1 - \ell H_{2}(-\lambda) - (1 - \ell) H_{2}(\lambda)}, 
\end{align}
and $H_{2}(-\lambda) = \partial_{\lambda} H_{1}(-\lambda)$. For $\Delta_{0} > 0$, the zero temperature energy is zero, which implies a zero distortion compression. The mean-field equations are
\begin{align}
&q_{0} = \ell  F_{1}( - \lambda) + (1 - \ell) F_{1}( \lambda) , \label{eq:q0_def}\\
&q_{1} = \ell F_{2}( - \lambda) + (1 - \ell) F_{2}( \lambda), \label{eq:q_def}\\
& \ell-\bar{\ell}  = \ell F_{1}( - \lambda) - (1 - \ell)F_{1}( \lambda), \label{eq:constraint}
\end{align}
where
\begin{align}
F_{k}( \lambda) &= \int Dz \left( 1 + \exp \left( \Theta(z, \lambda) \right) \right)^{-k},\\
\Theta(z, \lambda) & = - \frac{1}{ q_{0} - q_{1}}  \left[ \sqrt{\alpha q_{1}} z -  \frac{\alpha}{2} - \lambda \right],
\end{align}
and $Dz = dz \exp( - z^{2}/2)/\sqrt{2\pi}$. Here, Eq.\ref{eq:constraint} comes from the hard constraint on compression length.

We now explore various limits of the mean-field theory. 

\subsection{Phase Diagram}

Fig. \ref{fig:phase}A shows the zero temperature phase diagram obtained from the RS MFT describe above. In general, the order parameters will depend on the parameters $\alpha, \ell, \bar{\ell}$, but only through the following ratios $\alpha/\ell$, $C = \bar{\ell}/\ell$, and $\ell / (1 - \ell)$. We work exclusively within the replica symmetric ansatz. The phase boundaries are drawn in the following manner: if there exists a solution to the MFT Eqs. \ref{eq:qbar_mft} and \ref{eq:lambda_mft} that has $\delta>0$, we assume the system is in the lossy phase. The lossy phase is impossible if $\delta<0$, and we denote that by the condition (at zero temperature) that $\Delta_{0} = \Delta > 0$ (shaded blue in Fig. \ref{fig:phase}A). We have not logically ruled out the possibility that lossless compression happens outside this region.

For simplicity, we fix $\ell = 0.4$ to be able to visualize the phase diagram as $\alpha$ and $C$ are varied. For a given $\ell$, there is a maximal $\alpha^{*}$ above which lossless compression is impossible. This corresponds to the point where the solid black line meets the dashed black line, indicated by a star in Fig.\ref{fig:phase}A. We find numerically that $\alpha^{*}(\ell)$ is approximately quadratic in the interval $[0,1]$, and goes to zero at the boundaries. The peak obtains at $\ell \approx 0.53$, with $\alpha^{*}(\ell^{*}) \approx 0.4049$ (see Fig. \ref{fig:max_alpha} in the supplemental material).

\subsection{Small Lexicon Limit}

The limit $\alpha \to 0$ is somewhat trivial in our model, due to the explicit scaling we use for the Hamiltonian. Nevertheless, it is instructive to describe this limit. The RMFT yields
\begin{align}
q_{0} = \ell + \bar{\ell} - 2 \ell \bar{\ell}, \quad q_{1} = \ell - 2 \ell \bar{\ell} + \bar{\ell}^{2},
\end{align}
From which we get the order parameters
\begin{align}
r = \ell \bar{\ell}, \quad \Delta = \bar{\ell}( 1- \bar{\ell}).
\end{align}
Furthermore, in this limit, the minimum distortion is precisely zero. This is simply a consequence of the fact that every possible $\bar{c}$ will produce a distortion that scales like $P/N$, and thus tends to zero in the thermodynamic limit if $P = O(1)$.

\subsection{Large Embedding Dimension}

We can consider also the limit $\alpha \to \infty$, in which the embedding dimension becomes much larger than the size of the lexicon. In this limit, the embeddings become approximately orthogonal, and we expect to recover the weighted Hamming phase. This can be observed directly from the zero temperature RS MFT with $\Delta_{0} = 0$, in which we get
\begin{align}
q \approx \ell - \bar{\ell}, \quad D \sim \frac{\alpha (\ell - \bar{\ell})}{2}.
\end{align}
This follows from \ref{eq:min_hamming} by noting that $\langle w \rangle = P$ for random Gaussian embeddings.

\subsection{Signal Recovery Limit: $C = 1$}

In the limit that the compression ratio $C = 1$, our model is formally very close to the compressed sensing problem of signal reconstruction studied in \cite{ganguli2010statistical,vehkapera2013statistical}, except for the fact that our signal and message are binary variables. In this limit, the RS solution has positive entropy (unlike seemingly all solutions with $C<1$). For $\alpha/\ell \gtrapprox 1.3257 $, the only solution to Eqs. \ref{eq:qbar_mft} and \ref{eq:lambda_mft} is $q= 0$, which makes $D = 0$ and thus corresponds to perfect signal recovery. For smaller $\alpha/\ell$, the recovered signal will not be perfect. This regime falls inside the orange shaded region enclosed by the dashed black lines in Fig.\ref{fig:phase}A.

\subsection{Comparison to numerics}

It is important to note that in the zero temperature limit, the entropy is strictly negative throughout the phase diagram, except for compression ratios very close to one (inside the orange shaded region in Fig.\ref{fig:phase}A). On the surface, this would make our phase diagram meaningless. However, comparing to numerics, we find that the phase diagram is at the very least qualitatively useful, if not quantitatively correct. What it does not tell us is whether there is a sharp phase transition corresponding to a change in the replica overlap order parameter $\Delta$. Indeed, the numerics would suggest a continuous crossover in $\Delta$ (\ref{fig:phase}B).

\begin{figure*}[htbp!]

\includegraphics[width = \textwidth]{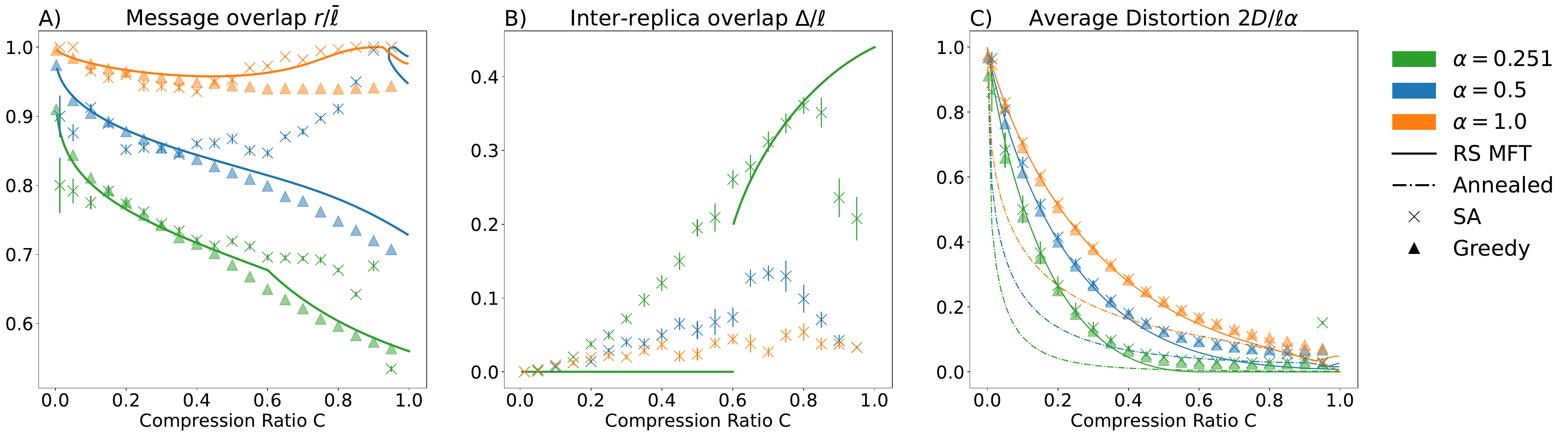}

	\caption{ {\it Numerics} Comparing numerical optimization via simulated annealing (SA) and a greedy algorithm (GA), with RS MFT. {\bf A)} The SA values (crosses) of the order parameters diverge from RS MFT prediction (solid curves) in the regime we expect to see the phase transition to lossless and abstractive compression (for the green points, past $C = 0.6$). There is also a striking deviation for the blue ($\alpha = 0.5$) for intermediate $C$, but agreement for small $C$ and $C \to 1$. Curiously, when disagreement between RS MFT and SA is large, the GA finds solutions with an order parameter that is very close to that predicted by the RS MFT. {\bf B)} The numerically computed inter-replica overlap order parameter shows a smooth rise and fall, in contrast to the theoretical prediction. Since the GA produces a unique minimizer for a given $c$, there is no comparison to be made here with SA, which generally finds multiple minimizers for a given quenched message.  {\bf C)} The average distortion is fairly well described by theory, with notable deviations for larger compression ratios, as in the previous plots. Evidently, the annealed lower bound is not very tight for small $C$, but gets better for increasing $C$. We used $N = 200$ and average over $10$ embedding (disorder) realizations. The order parameters are computed by estimating the low energy spectrum of the Hamiltonian and computing a truncated temperature average (see Appendix \ref{sec:numerics} for details)}
	\label{fig:numerics}
\end{figure*}

In Fig. \ref{fig:numerics}, we compare the theoretical results to two optimization algorithms: simulated annealing (SA) and a greedy algorithm (GA). It is quite interesting to note that in most regions of the phase diagram, the greedy algorithm is significantly faster and performs nearly as well as, if not better than, SA in minimizing distortion. Perhaps the most interesting observation is that in the paraphrastic phase, the overlap order parameter obtained via GA is well approximated by the RS MFT. Overall, we observe that the RS MFT offers decent predictions for large $\alpha$ (approaching the Hamming phase), and for small compression ratio $C$. However, approaching the region with $\Delta > 0$, the theory apparently breaks down. We attribute this to both the patently wrong approximation of replica symmetry and finite-size effects in the numerical simulations (which we performed for rather small systems). 

\section{Discussion}

We have introduced a statistical mechanics model of semantic compression, and solved it under a replica symmetric ansatz. In this work, we have learned that even with completely random embeddings, semantic compression undergoes a transition between unique compression and paraphrastic compression. The detailed structure of the phase diagram is governed both by properties of the semantic space (relative embedding dimension $\alpha$), as well as the compression ratio. We have also found crossover behavior between extractive and abstractive summarization. 

While this work has focused on formulating the mathematical problem and solving it in the mean field limit, it raises tantalizing questions about fitting to real-world language data. For instance, where does typical communication lie on this phase diagram? We may speculate using some details from modern language models. Typical lexicon size is of order $10^{4}$, while embedding dimensions are $P \sim O(10^{2})$, giving $\alpha \sim  0.01$. At this value, the compression ratio does not have to be very large before reaching the phase boundary. Across this phase boundary, there will tend to be many summaries which are very good. In other words, for any given message $c$, it will be possible to find many paraphrases which roughly mean the same thing. This is, at least intuitively, precisely the situation with natural language. Mathematically, this means the probability distribution of a message conditioned on its meaning, $P(c |M)$, has nonzero entropy. A similar quantity, referred to as the ``wording information", was recently shown to be nonzero using large language models \cite{sivan2024information}. Estimating $\alpha$ is more difficult for humans. We may estimate a lexicon size which is roughly comparable in order of magnitude \cite{brysbaert2016many}, but it is anyone's guess what the dimension of a human's semantic space may be. For instance, the fMRI study of \cite{Huth2012} found that the semantic space should have at least four dimensions, setting a very small lower bound on $\alpha$. Other sources seem comfortable with semantic (or conceptual) spaces with a huge number of dimensions, resulting in a large relative embedding dimension \cite{Piantadosi2024}. Of course, the structure of real semantic embeddings is not random, and the joint embedding of a message is likely not additive but context dependent. It would be interesting to explore how these properties influence the compressibility of a language. For instance, structured embeddings might allow for meaning-preserving compression even at very large $\alpha$. We leave these exciting questions for future work.

{\it Optimization and Mixed Integer Linear Programming}

The process of semantic compression we have studied in this paper is an example of a mixed-integer linear programming problem. This tells us that the problem is in fact NP-hard. The remarkable fact is that we can find a fast algorithm which is $O(\bar{T} L)$ that can give a good solution to this problem. This is not terribly surprising, since our formulation of semantic compression is very similar to the partition problem, which has been called the ``easiest" NP-hard problem \cite{Korf2009,hayes2002computing}. We showed that a simple greedy algorithm often gives a very low distortion. Furthermore, we showed numerically that the greedy algorithm gives a solution that is well described by the replica symmetric order parameters.

{\it Communication and an Alignment Problem:} Communication is rife with misunderstanding, and our theory actually sheds some light on a potential mechanism. Suppose Alice sends a message $c_{A}$ to Bob, who then produces a compressed message $\bar{c}_{B}$ which, according to Bob, is distortion minimizing. Bob chooses his compression using his personal semantic embeddings. Now Alice can compare $\bar{c}_{B}$ to $c_{A}$, by either computing the overlap or the distortion. But to calculate distortion, Alice can only use her own embeddings. If Alice and Bob have perfectly aligned semantic embeddings, then the message $\bar{c}_{B}$ would presumably look like a compression that Alice could have come up with. Therefore, she will agree that $\bar{c}_{B}$ means the same thing as $c_{A}$, and will conclude that Bob understood the original message. However, if the embeddings are not aligned, the distortion will increase with the degree of misalignment. Therefore, the simple fact that Alice recognizes semantic similarity between her original message, and Bob's repeated version of her message, implies some degree of alignment between their separate, private, semantic spaces. Note also that the crucial thing is not that they both have vectors pointing in the same direction - if all of Bob's embeddings are related to Alice's embeddings by the same orthogonal transformation, then although their embedding vectors might be different, the Hamiltonian, and hence distortion function, is unchanged. So really, the crucial property is the relative positions of the embeddings. Surprisingly, there is experimental support for shared semantic dimensions \cite{Huth2012} and structure in neural activity \cite{chen2017shared}. But after thinking through the problem of communication, we conclude that such a shared space is hardly surprising at all, and in fact is necessary for people speaking a common language.

In fact, the thought experiment above illustrates a general {\it semantic alignment problem}: does communication between individuals require an alignment between their semantic spaces? In the most general setting, we might replace alignment with isomorphism, especially if the semantic spaces are mathematically different spaces. For the model of semantic compression we consider in this paper, the notion of alignment is taken from linear algebra. However, this formulation extends beyond the realm of human communication. Suppose Bob is a large language model (LLM), which are known to be excellent summarizers \footnote{The proof of the pudding is in the eating: when prompted with the request to ``Summarize the following text in one sentence: " and then given the previous paragraph, OpenAI's GPT-4 produced the following: ``Miscommunication arises from misaligned semantic embeddings between individuals, necessitating a shared semantic space for effective communication."}. The scenario above resembles modern variants of Turing's imitation game \cite{turing2021computing}, wherein Bob simply needs to convince Alice that he is human. In this semantic compression game, Bob simply needs to convince Alice that he {\it understands} her, by doing what any good student does: summarizing what Alice says in his own words, but without losing the meaning of the original message. And if Bob is an LLM, then in all likelihood he will win this game. We conclude that for this to be possible, there must exist an isomorphism between the semantic space constructed by our brains, and the latent representations utilized by LLMs in performing their computation. We speculate that this mathematical isomorphism between semantic spaces is at the root of what is referred to as ``common ground" in linguistics and philosophy, which encompasses the knowledge base shared between individuals that provides the scaffolding for effective communication \cite{hao2023common}.

\smallskip
{\bf In summary}, we have formulated semantic compression as a combinatorial optimization problem for a spin glass Hamiltonian, and solved the statistical mechanics in the replica symmetric limit. We find that for a lexicon randomly embedded in semantic space, the compressibility of a message undergoes both phase transitions and crossover behavior, as a function of embedding dimension, message length, and compression ratio. For sufficiently small embedding dimension, small compression ratios tend to incur distortion, whereas larger compression ratios are lossless. Furthermore, the compressed messages in the lossless phase tend to be abstractive, using lexical items outside the original message for more efficient summaries. For larger embedding dimension, there is no phase transition, and compression always incurs a cost. In this region of the phase diagram, the compressed messages tend to be extractive, restricted to using lexical items from the original message. Finally, we show that while the original optimization problem falls in the class of mixed-integer linear programming, and is therefore NP-hard, we were able to find an efficient greedy algorithm that is competitive with the more costly simulated annealing. 

We assumed semantic embeddings had no correlations, which probably does not reflect the structure of such embeddings in the wild. We will examine the influence of structure and correlation in semantic space in future work. Our theory also has introduced two novel order parameters, the overlap between original and compression, and the inter-replica overlap, which can be applied to the study of language. We hope to explore this in follow-up work. \\

\vspace{20pt}

\noindent {\bf Acknowledgements} I have benefited from numerous conversations on this topic, and would like to thank Mikhail Katkov, Misha Tsodyks, Weishun Zhong for their feedback. I am especially indebted to Bruno Loureiro and Francesca Mignacco for some critical insights at the initial stages. I acknowledge the support of the Eric and Wendy Schmidt Membership in Biology, the Simons Foundation, and the Starr Foundation Member Fund in Biology at the Institute for Advanced Study, where much of this work was completed.

\bibliography{Semantics.bib}


\appendix

\begin{widetext}

\section{Numerical Simulations}\label{sec:numerics}

For the numerics in Fig.\ref{fig:numerics}, we searched for compressions with simulated annealing \cite{Parisi1987}. For a given quenched message $c$, we initialized the search by taking a random subset of $\bar{L}$ entries which were equal to one, and setting the rest of the entries to zero. This provides $\bar{c}_{0}$. A stochastic update is made on a state $\bar{c}_{t}$ by randomly flipping two bits from $0 \to 1$ and $1 \to 0$, to give a new state $\bar{c}_{t}^{*}$ with the same norm. The update rule for the compression $\bar{c}_{t}$ is probabilistic, and given by

\begin{align}
\bar{c}_{t+1} = \begin{cases}
\bar{c}_{t}^{*}, \quad {\rm w/ \quad probability}  \quad P_{t}\\
\bar{c}_{t}, \quad {\rm w/ \quad probability}  \quad  1 - P_{t}
\end{cases},
 \quad P_{t} = \exp \left( - \gamma_{t} ( H(c, \bar{c}_{t}^{*})- H(c, \bar{c}_{t}))\right).
\end{align}
Here, $H(c, c')$ is the Hamiltonian defined in the Eq. \ref{eq:H}, and $\gamma_{t}$ is the inverse temperature of the Monte Carlo algorithm, which we update according to the annealing schedule:

\begin{align}
\gamma_{t+1} = \begin{cases}
r \gamma_{t}, \quad {\rm if} \quad H(c, \bar{c}_{t+1}) < H(c, \bar{c}_{t})\\
\gamma_{t}, \quad {\rm otherwise}
\end{cases}
\end{align}

For our simulations, we used $\gamma_{0} = 0.04$, and $r = 1.01$.  After a fixed number of steps (we chose $3000$), we stop the algorithm. Running this many times gives an ensemble of $\bar{c}_{a}$ from we define an empirical probability with a given $\beta$ that we will eventually take to be large:

\begin{align}
p_{e}(\bar{c}_{a}) = \frac{e^{ - \beta H(\bar{c}_{a})}}{\sum_{b} e^{ - \beta H(\bar{c}_{b})}} . 
\end{align}

We suppress the $c$ argument of the Hamiltonian, since this variable is fixed throughout. With this empirical measure, we compute the order parameters
\begin{align}
\hat{q}_{0}(\beta) &= \sum_{a} p_{e}(\bar{c}_{a}) \frac{1}{N} ||c - \bar{c}_{a}||_{1}, \quad \hat{q}_{1}(\beta) = \sum_{a, b} p_{e}(\bar{c}_{a}) p_{e}(\bar{c}_{b})   \frac{1}{N}(c - \bar{c}_{a}) \cdot (c - \bar{c}_{b}).
\end{align}

From these we get the empirical overlap and EA order parameter using 

\begin{align}
r_{e} = \frac{1}{2} ( \ell + \bar{\ell} - \hat{q}_{0}), \quad  \Delta_{e}= \hat{q}_{0} - \hat{q}_{1}. 
\end{align}

For Fig.\ref{fig:numerics}, we plotted the numerical order parameters using $\beta = 10$.

\subsection{Greedy Algorithm}

The greedy algorithm discussed in the main text proceeds as follows: define the initial vector as the message vector embedding

\begin{align}
X_{0} \equiv X(c) = \sum_{i = 1}^{N} c_{i} E_{i}.
\end{align}
Next, find the lexical item closest to $X_{0}$, 

\begin{align}
i_{1} = \underset{k}{\rm argmax} || X_{0} - E_{k}||^{2}.
\end{align}

Then, update the vector by subtracting this closest lexical item:

\begin{align}
X_{1} = X_{0} - E_{i_{1}}.
\end{align}

For a general step, the updating of the vector is

\begin{align}
i_{t} &= \underset{k}{\mathrm{argmin}} || X_{t-1} - E_{k}||^{2}, \quad 
X_{t} = X_{t-1} - E_{i_{t}}
\end{align}

Finally, to get the desired length of compression, terminate the process after finding $i_{\bar{L}}$. The compressed message then has $\bar{c}_{i_{t}} = 1$ for $t =1, ..., \bar{L}$. 

For Fig. \ref{fig:numerics}, we employ the greedy algorithm for systems with $L = 1000$, and obtain all the curves with quenched embeddings, averaging over 500 different random target messages. 

\section{Replica Theory for Semantic Compression}

The energy function is given by the Euclidean square distance between the original phrase embedding $X(c)$ and the paraphrase $X(\bar{c})$ .

\begin{align}
H(c, \bar{c}) = \frac{1}{2   N} || X(c) - X(\bar{c})||^{2} = \frac{1}{2 N } \sum_{i, j = 1}^{N} (c_{i} - \bar{c}_{i}) E_{i} \cdot E_{j} ( c_{j} - \bar{c}_{j}) \equiv \frac{1}{ 2 N} \sum_{\mu = 1}^{P} \left( \sigma \cdot E^{\mu}\right)^{2}, \quad \sigma = c - \bar{c}.
\end{align}

The partition function that we compute will include the constraint that $||\bar{c}||_{1} = \bar{L}$, i.e. the one-norm of the paraphrase configuration is fixed. Furthermore, in this problem we are treating the original phrase $c$ and the embeddings $E$ as quenched variables. The partition function at temperature $T = \beta^{-1}$ is then

\begin{align}
Z_{\beta}(c, E, \bar{L}) = \sum_{\bar{c}_{i}} e^{ - \beta H } \delta( || \bar{c}||_{1} - \bar{L})  = \sum_{\bar{c}} \int d\bar{x} \exp\left[ - \frac{\beta}{2 N} \sum_{\mu = 1}^{P} ( \sigma \cdot E^{\mu})^{2} + i \bar{x} \left(\bar{L}- \sum_{i} \bar{c}_{i} \right) \right].
\end{align}
The free energy density which follows from partition function is

\begin{align}
f_{\beta}(c, E, \bar{L})  = - \frac{1}{\beta N} \log Z_{\beta}(c, E, \bar{L}) .
\end{align}

It is not immediately obvious that this quantity should be self-averaging, and we could in principle consider the distribution of the free energy density. But for simplicity, we take the mean value. 

\begin{align}
f_{\beta} (L, \bar{L}, \theta) =  - \frac{1}{\beta N}  \frac{\sum_{c} \delta(||c||_{1} - L)}{\mathcal{N}(L)} \int dP(E) \log Z_{\beta} (c, E, \bar{L}) \equiv - \frac{1}{\beta N} \langle \log Z_{\beta}(c, E, \bar{L})\rangle_{c, E}  ,
\end{align}

where $\mathcal{N}(L) = \sum_{c} \delta( || c||_{1} - L)$ is a normalization factor for the distribution of $c$ which counts the number of the distinct phrases of length $L$. We also denote by $\theta$ the set of all parameters that characterize the distribution over embeddings $E$. We assume this expression for the free energy follows from the zero replica limit

\begin{align}
f_{\beta}(L, \bar{L}, \theta) = - \frac{1}{\beta N} \lim_{n \to 0} \frac{1}{n} \left(\langle Z_{\beta}^{n}(c, E, \bar{L}) \rangle_{c, E} - 1 \right).
\end{align}

From the free energy, we are interested in the ground state energy, which in our problem has the interpretation of the minimal distortion paraphrase per length of total lexicon. This is given by the zero temperature limit

\begin{align}
D_{min} = \lim_{\beta \to \infty} f_{\beta}.
\end{align}
This can be converted easily to a more reasonable measure which is the distortion per dimension of the embedding by dividing the RHS by $\alpha$. Another possible measure is the distortion per length of original phrase, which requires just dividing the RHS by $\ell = L/N$. Another interesting quantity is the average energy density (distortion)

\begin{align}
D = \frac{1}{N}\langle H \rangle = \partial_{\beta} \left( \beta f_{\beta}\right)
\end{align}

\subsection{Replica Partition function}

Here we calculate the quenched averaging of the replicated partition function. First, the replicated partition function takes the form

\begin{align}
Z_{\beta}^{n}(c, E) &=  \sum_{\bar{c}_{i}^{a}} \int d\bar{x}^{a} \exp\left[ - \frac{\beta}{2 N} \sum_{a} \sum_{\mu = 1}^{P} ( \sigma^{a} \cdot E^{\mu})^{2} + i \bar{x}^{a} \left( \sum_{i} \bar{c}_{i}^{a} - \bar{L}\right) \right]\\
& =  \sum_{\bar{c}_{i}^{a}} \int \prod_{a} d\bar{x}^{a} \prod_{\mu} Du_{\mu}^{a} \exp\left[ i \sqrt{\frac{\beta}{ N}} \sum_{\mu, a} u_{\mu}^{a}( \sigma^{a} \cdot E^{\mu}) + i \bar{x}^{a} \left( \bar{L}-\sum_{i} \bar{c}_{i}^{a} \right) \right],
\end{align}

where we have linearized the argument of the exponential using a Hubbard-Stratonovich transformation, and have introduced for notational shorthand the Gaussian measure $Du = d^{ - u^{2}/2}/\sqrt{2\pi}$ such that $\int Du = 1$. We assume the embeddings are drawn randomly i.i.d. from a non-centered Gaussian

\begin{align}
\langle E_{i}^{\mu} \rangle = b^{\mu} , \quad \langle \left(E_{i}^{\mu}\right)^{2} \rangle_{c} = \sigma^{2}.
\end{align}

After averaging the replica partition function over $E$, and defining our order parameter

\begin{align}
q_{ab} = \frac{1}{N} \sum_{i} \sigma_{i}^{a} \sigma_{i}^{b}. \label{eq:q_order}
\end{align}

We get the averaged partition function
\begin{align}
\langle Z_{\beta}^{n}(c, E) \rangle_{E} & =  \sum_{\bar{c}_{i}^{a}} \int \prod_{a} d\bar{x}^{a} \prod_{\mu} Du_{\mu}^{a} \exp\left[ i \sqrt{\frac{\beta}{ N}} \sum_{\mu, a} u_{\mu}^{a}( \sum_{i} \sigma^{a}_{i}) b^{\mu} - \frac{\beta \sigma^{2} }{2 } \sum_{\mu} \sum_{a,b} u_{\mu}^{a} u_{\mu}^{b}  q_{ab} + i \bar{x}^{a} \left( \bar{L}- \sum_{i} \bar{c}_{i}^{a} \right) \right].
\end{align}

There are a few simplifications to be made at the present time. First, we use the fact that the constrains on the $L_{1}$ norm of the $c$ and $\bar{c}$ mean that

\begin{align}
\sum_{i} \sigma_{i}^{a} = ||c||_{1} - ||\bar{c}^{a}||_{1} = L - \bar{L} = N m , \quad m \equiv  \ell - \bar{\ell},
\end{align}

where we have introduced a parameter $m$ which measures the difference in magnetization density of each configuration. Integrating the $u_{\mu}^{a}$ we end up with

\begin{align}
\langle Z_{\beta}^{n}(c, E) \rangle_{E} & =  \sum_{\bar{c}_{i}^{a}} \int \prod_{a} d\bar{x}^{a}  \exp\left[ - \frac{P}{2} \log \det \left( 1 + \beta \sigma^{2} q \right) - \frac{N \beta m^{2}  \left( \sum_{\mu} b_{\mu}^{2} \right)}{ 2 } {\bf l}^{T} K {\bf l} + i \bar{x}^{a} \left(  \bar{L}-\sum_{i} \bar{c}_{i}^{a} \right) \right],
\end{align}

where $K = \left(1 + \beta \sigma^{2}  q \right)^{-1}$, and ${\bf l} \in \mathbbm{R}^{n}$ with ${\bf l}_{a} = 1$. Next, we seek to reduce the redundancy in model parameters. Defining

\begin{align}
\mu = \frac{\sum_{\mu} b_{\mu}^{2}}{\sigma^{2}},
\end{align}

and redefining $\beta \sigma^{2} \to  \beta$, we get
\begin{align}
\langle Z_{\beta}^{n}(c, E) \rangle_{E} & =  \sum_{\bar{c}_{i}^{a}} \int \prod_{a} d\bar{x}^{a}  \exp\left[ - N \frac{\alpha }{2} \log \det \left( 1 + \beta q \right) - \frac{N \beta \mu m^{2} }{ 2 } {\bf l}^{T} K {\bf l} + i \bar{x}^{a} \left(\bar{L}- \sum_{i} \bar{c}_{i}^{a} \right) \right].
\end{align}

In order to now average over configurations $\bar{c}$, we introduce the Lagrange multiplier $\hat{q}$ to enforce the constraint \ref{eq:q_order}. This will render the partition function

\begin{align}
\langle Z_{\beta}^{n}(c, E) \rangle_{E}  = \int  d [x, \hat{q}, q] \, \exp & \left[  i N \sum_{a \le b} \hat{q}_{ab} q_{ab}  - N \frac{\alpha }{2} \log \det \left( 1 +\beta q \right) - \frac{N \beta\mu m^{2} }{ 2 } {\bf l}^{T} K {\bf l} + i \bar{L} \sum_{a} \bar{x}^{a}  + \sum_{i} \log \mathcal{Z}[\hat{q}, \bar{x}, c_{i}] \right],
\end{align}
where we have denoted the integration measure

\begin{align}
d[x,  \hat{q}, q] = \prod_{a} dx^{a} \prod_{a \le b} d \hat{q}_{ab} dq_{ab},
\end{align}
and we have introduced
\begin{align}
\mathcal{Z}[ \hat{q}_{ab}, \bar{x}^{a}, c_{i}] = \sum_{\bar{c}^{a}} \exp \left( - i  \sum_{a \le b} \hat{q}_{ab} (c_{i} - \bar{c}^{a}) ( c_{i} - \bar{c}^{b}) - i \bar{x}^{a} \bar{c}^{a} \right).
\end{align}

Now note that since $c_{i}$ can only take on two values, and this single-site partition function does not otherwise depend on the site index, it similarity will only take on two values: $\mathcal{Z}[ \hat{q}, \bar{x}, 0]$ and $\mathcal{Z}[ \hat{q}, \bar{x}, 1]$. Furthermore, since the norm of $c$ is imposed to be $L$, we have that

\begin{align}
\prod_{i} \mathcal{Z}[ \hat{q}_{ab}, \bar{x}^{a}, c_{i}] = \left(\mathcal{Z}[ \hat{q}_{ab}, \bar{x}^{a}, 1]\right)^{L} \left(\mathcal{Z}[ \hat{q}_{ab}, \bar{x}^{a}, 0]\right)^{N-L}.
\end{align}

Therefore, the partition function does not depend on the detailed configuration of $c$, but only on its total length $L$. This means we can trivially take the average over $c$ and get

\begin{align}
\langle Z_{\beta}^{n}(c, E) \rangle_{c, E} =   \int  d [x, \hat{q}, q]  \, \exp & \Big[  i N \sum_{a \le b} \hat{q}_{ab} q_{ab}  - N \frac{\alpha }{2} \log \det \left( 1 +\beta q \right) - \frac{N \beta\mu m^{2} }{ 2 } {\bf l}^{T} K {\bf l} + i \bar{L} \sum_{a} \bar{x}^{a} \\
&  + L \log \mathcal{Z}[\hat{q}, \bar{x}, 1]  + (N - L) \log \mathcal{Z}[\hat{q}, \bar{x}, 0] \Big].
\end{align}

Next, we find explicit expressions for $\mathcal{Z}$ that we can work with. Defining

\begin{align}
\mathcal{Z}[\hat{q}, \bar{x}] \equiv \mathcal{Z}[ \hat{q}_{ab}, \bar{x}^{a},0]   = \sum_{\bar{c}^{a}} \exp \left( - i  \sum_{a \le b} \hat{q}_{ab} \bar{c}^{a}  \bar{c}^{b} - i \sum_{a} \bar{x}^{a} \bar{c}^{a} \right),
\end{align}
we find that 

\begin{align}
\mathcal{Z}[ \hat{q}_{ab}, \bar{x}^{a},1]  = e^{ - i \sum_{a} \bar{x}^{a} } \mathcal{Z}\left[ \hat{q}, - \bar{x} \right]. \\
\end{align}

With this, we get the replica partition function in a form which will allow us to perform a saddle-point calculation

\begin{align}
\langle Z_{\beta}^{n}(c, E) \rangle_{c, E} &  = \int d [x, \hat{q}, q]  \exp\left( N \mathcal{H}_{n}\right),
\end{align}

where
\begin{align}
\mathcal{H}_{n} & = - \frac{\alpha}{2} \log \det \left(1 + \beta q\right) - \frac{\beta \mu m^{2}}{2} \ell^{T} K \ell +  \sum_{a \le b} Q_{ab} q_{ab} - \sum_{a} X^{a} m + t \log \mathcal{Z}[Q, - X] + (1 - t) \log \mathcal{Z}[ Q,X], \\
K &= \left(1 + \beta q\right)^{-1}, \quad \ell_{i} = 1, \quad  m = \ell - \bar{\ell}, \quad Q \equiv  i \hat{q}, \quad X \equiv  i \bar{x},
\end{align}
and the ``single site partition function" is
\begin{align}
\mathcal{Z}[Q,X] = \sum_{s^{a} \in \{0,1\}} e^{ -  \sum_{a \le b}Q_{ab}s^{a} s^{b} -  \sum_{a} X^{a} s^{a}}.
\end{align}
Now defining for some function of spins $G(s)$, the single site correlation function is defined as
\begin{align}
\langle G(s) \rangle_{\pm  } \equiv  \frac{1}{\mathcal{Z}[Q, \pm X] }\sum_{s^{a} \in \{0,1\}} G(s) e^{ -  \sum_{a \le b}Q_{ab}s^{a} s^{b}   -   \sum_{a} ( \pm X^{a}) s^{a} }.
\end{align}

The saddle-point equations are

\begin{align}
\frac{\delta}{\delta Q_{ab}} \mathcal{H}_{n} &=  q_{ab}  -  \ell \langle s^{a} s^{b}\rangle_{-} -  (1 - \ell) \langle s^{a} s^{b}\rangle_{+},\\
\frac{\delta}{\delta X_{a}} \mathcal{H}_{n} & =- m +  \ell \langle s^{a} \rangle_{-} -  (1 - \ell) \langle s^{a} \rangle_{+},\\
\frac{\delta }{\delta q_{ab}} \mathcal{H}_{n} & = - \alpha \beta K_{ab} +   \beta^{2} \mu m^{2} ( K {\bf l})_{a} ( K {\bf l})_{b} + Q_{ab}, \quad a \ne b,\\
\frac{\delta }{\delta q_{aa}} \mathcal{H}_{n} & = - \frac{\alpha \beta }{2}   K_{ab} +  \frac{ \beta^{2} \mu m^{2}}{2} ( K {\bf l})_{a} ( K {\bf l})_{b} + Q_{aa}.
\end{align}

\subsection{Replica symmetric ansatz}

Assuming replica symmetric order parameters
\begin{align}
q_{ab} = (q_{0} - q) \delta_{ab} + q, \quad Q_{ab} = (Q_{0} - Q) \delta_{ab} + Q, \quad X^{a} = X, 
\end{align}

implies also that
\begin{align}
K_{ab} = (K_{0} - K) \delta_{ab} + K.
\end{align}

We label $\Delta = q_{0} - q$, since this appears very often below. Then the components of the inverse propagator $K$ are

\begin{align}
K_{0} - K &=  \frac{1}{1 + \beta \Delta}, \quad K  = - \frac{\beta q}{(1 + \beta \Delta) ( 1 + \beta \Delta + n \beta q)}.
\end{align}

We first evaluate the single-site partition function

\begin{align}
\mathcal{Z}[ Q, X] &= \sum_{s^{a}} \exp \left[ - ( Q_{0} - \frac{1}{2} Q + X) \sum_{a} s_{a} - \frac{1}{2} Q \left(\sum_{a} s_{a}\right)^{2} \right],\\
& = \int Dz \sum_{s^{a}} e^{ \sqrt{ - Q} z \sum_{a} s_{a} -  ( Q_{0} - \frac{1}{2} Q + X) \sum_{a} s_{a}} ,\\
& = \int Dz \left( 1 + e^{- \Theta_{+}(z)}\right)^{n}, \quad \Theta_{\pm} (z) = -\sqrt{ - Q}  z  +  Q_{0} - \frac{1}{2} Q \pm  X ,\\
& \to 1 + n \int Dz \log \left(1 + e^{ - \Theta_{+}(z)}\right) + O(n^{2}).
\end{align}

On the replica symmetric saddle, $\mathcal{H}_{n}^{*} =  n \mathcal{H}_{0}^{*} + O(n^{2})$, where

\begin{align}
\mathcal{H}_{0}^{*}&= - \frac{\alpha}{2} \left[ \log(1 + \beta \Delta) + \frac{\beta q}{1 + \beta \Delta} \right] - \frac{\beta \mu m^{2}}{2} \frac{1}{1 + \beta \Delta} +  Q_{0} q_{0} - \frac{1}{2}Q q - X m \\
& + \ell  \int Dz \log \left(1 + e^{ - \Theta_{-}(z)}\right) + (1 - \ell) \int Dz \log \left(1 + e^{ - \Theta_{+}(z)}\right) + O(n^{2}),
\end{align}
so that the free energy density becomes a function of the relative message and compression lengths, $\ell$ and $\bar{\ell}$, respectively, as well as the relative embedding dimension $\alpha = P/N$, and the average mean of the embedding vectors $\mu$: 

\begin{align}
f_{\beta}(\ell, \bar{\ell}, \alpha, \mu) &= - \frac{1}{\beta} \mathcal{H}_{0}^{*} =  \frac{\alpha}{2 \beta} \left[ \log(1 + \beta \Delta) + \frac{\beta q}{1 + \beta \Delta} \right] + \frac{ \mu m^{2}}{2} \frac{1}{1 + \beta \Delta} -  \frac{1}{\beta} Q_{0} q_{0} + \frac{1}{2 \beta}Q q + \frac{1}{\beta} X m \\
& - \frac{1}{\beta} \ell  \int Dz \log \left(1 + e^{ - \Theta_{-}(z)}\right) - \frac{1}{\beta} (1 - \ell) \int Dz \log \left(1 + e^{ - \Theta_{+}(z)}\right) + O(n^{2}).
\end{align}

The saddle-point equations become

\begin{align}
Q_{0} &= \frac{\alpha \beta }{2} \left[ \frac{1}{1 + \beta \Delta} - \frac{\beta q}{(1 + \beta \Delta)^{2}}\right] - \frac{\beta^{2} \mu m^{2}}{2 (1 + \beta \Delta)^{2}},\\
- Q & = \frac{\beta^{2} ( \alpha q + \mu m^{2})}{(1 + \beta \Delta)^{2}} , \\
\ell - \bar{\ell} & =  \ell \int Dz \frac{1}{1 + e^{\Theta_{-}(z)}} - (1 - \ell) \int Dz \frac{1}{1 + e^{ \Theta_{+}(z)}} , \\
q_{0} &= \ell \int Dz \frac{1}{1 + e^{\Theta_{-}(z)}} + (1 - \ell) \int Dz \frac{1}{1 + e^{ \Theta_{+}(z)}}, \\
q &= \ell \int Dz \frac{1}{\left(1 + e^{\Theta_{-}(z)} \right)^{2}} + (1 - \ell) \int Dz \frac{1}{\left(1 + e^{ \Theta_{+}(z)}\right)^{2}}.
\end{align}

Plugging these into the free energy density affords some simplifications:

\begin{align}
f_{\beta}(\ell, \bar{\ell}, \alpha, \mu) 
& =  \frac{\alpha }{2 \beta } \log(1 + \beta \Delta) -  \frac{\alpha  \Delta}{2  (1 + \beta \Delta)^{2}}    \left( 1 + \beta  \Delta - \beta q  \right)  +  \frac{ \mu m^{2}}{2} \frac{(1 + 2 \beta  \Delta)}{ (1 + \beta \Delta)^{2}} \\
&+ \frac{X  (\ell - \bar{\ell})}{\beta} - \frac{\ell}{\beta}  \int Dz \log \left(1 + e^{ - \Theta_{-}(z)}\right) - \frac{(1 - \ell)}{\beta} \int Dz \log \left(1 + e^{ - \Theta_{+}(z)}\right) .
\end{align}

{\bf Summary of Replica Symmetric Mean Field Theory:}  The self-consistent mean-field equations are

\begin{align}
\ell - \bar{\ell} & =  \ell \int Dz \frac{1}{1 + e^{\Theta_{-}(z)}} - (1 - \ell) \int Dz \frac{1}{1 + e^{ \Theta_{+}(z)}} , \\
q_{0} &= \ell \int Dz \frac{1}{1 + e^{\Theta_{-}(z)}} + (1 - \ell) \int Dz \frac{1}{1 + e^{ \Theta_{+}(z)}}, \\
q &= \ell \int Dz \frac{1}{\left(1 + e^{\Theta_{-}(z)} \right)^{2}} + (1 - \ell) \int Dz \frac{1}{\left(1 + e^{ \Theta_{+}(z)}\right)^{2}}.
\end{align}

where
\begin{align}
\Theta_{\pm}(z) = - \frac{\beta}{1 + \beta \Delta} \sqrt{ \alpha q + \mu m^{2}} z + \frac{\alpha \beta}{2 (1 + \beta \Delta)} \pm X,
\end{align}

and the free energy density is

\begin{align}
f& =  \frac{\alpha }{2 \beta } \log(1 + \beta \Delta) -  \frac{\alpha  \Delta}{2  (1 + \beta \Delta)^{2}}    \left( 1 + \beta  \Delta - \beta q  \right)  +  \frac{ \mu m^{2}}{2} \frac{(1 + 2 \beta  \Delta)}{ (1 + \beta \Delta)^{2}} \\
&+ \frac{X (\ell - \bar{\ell})}{\beta} - \frac{\ell}{\beta}  \int Dz \log \left(1 + e^{ - \Theta_{-}(z)}\right) - \frac{(1 - \ell)}{\beta} \int Dz \log \left(1 + e^{ - \Theta_{+}(z)}\right) .
\end{align}

Differentiating the free energy density also gives the average distortion quoted in the main text:

\begin{align}
D(T)  = \frac{1}{2 (1 + \beta \Delta)^{2}} \left[ \alpha ( q_{0} + \beta \Delta^{2})+ \mu m^{2} \right].
\end{align}

\subsection{Zero temperature limits}

\subsubsection{Finite Variance Limit}

If we take $q_{0} - q = \Delta$ to be finite in the zero temperature limit, the RS MFT reduces to

\subsubsection{Zero Variance Lossy limit}
In the zero temperature limit, we assume 

\begin{align}
q_{0} - q  =  T \delta   + O(T^{2}),
\end{align}

and we change variables

\begin{align}
X \equiv \frac{ \beta \lambda}{1 + \beta \Delta} ,
\end{align}

which allows us to write
\begin{align}
\Theta_{\pm }(z) &= \frac{\beta}{1 + \beta \Delta} \left[ - \sqrt{\alpha q + \mu m^{2}} z + \frac{\alpha}{2} \pm  \lambda \right] = \frac{1}{T} \left[ - A_{0} z + B (\pm \lambda)\right] - A_{1} z + O(T^{2}),\\
A_{0} &= \frac{\sqrt{\alpha \bar{q} + \mu m^{2}}}{1 + \delta} , \quad B(\lambda) = \frac{\alpha/2 +  \lambda}{1 + \delta}.
\end{align}

Note that without the hard constraint on the compression length, we would not be permitted to make such a change. This change of variables ends up being very useful in finding a simple expression for $\delta$. Before we find this expression, we first show that $\Delta = O(T)$. Using the MFT to write $\Delta$, 

\begin{align}
\Delta = q_{0} - q = \ell \int Dz \frac{e^{ \Theta_{-} (z)}}{(1 + e^{ \Theta_{-}(z)})^{2}} + (1-\ell) \int Dz \frac{e^{ \Theta_{+}(z)}}{(1 + e^{ \Theta_{+}(z)})^{2}} .
\end{align}

Now changing variables $z = (B(\lambda) - T \theta)/A_{0}$, and neglecting the extra $T$ dependence in $\Theta$, we get

\begin{align}
\int Dz \frac{e^{ \Theta(z)}}{(1 + e^{ \Theta(z)})^{2}} &= \int \frac{1}{\sqrt{2\pi}} \frac{T d\theta}{A_{0}} e^{ - \frac{1}{2A_{0}^{2}} ( T \theta - B(\lambda))^{2}} \frac{e^{\theta}}{(1 + e^{\theta})^{2}}  = T \frac{1}{\sqrt{2\pi} A_{0}} e^{ - \frac{B(\lambda)^{2}}{2 A_{0}^{2}}} \int d\theta \left( - \partial_{\theta}  \right) \frac{1}{1 + e^{\theta}} + O(T^{2})\\
& = T \frac{1}{\sqrt{2\pi} A_{0}} e^{ - \frac{B(\lambda)^{2}}{2 A_{0}^{2}}}  + O(T^{2})
\end{align}

Expanding $\Delta$ to order $T$ then gives

\begin{align}
\delta &= \frac{\ell}{\sqrt{2\pi} A_{0}} e^{ - \frac{B(-\lambda)^{2}}{2 A_{0}^{2}}}  + \frac{(1 - \ell)}{\sqrt{2\pi} A_{0}} e^{ - \frac{B(\lambda)^{2}}{2 A_{0}^{2}}} 
 = \ell (1 + \delta) H_{2}(-\lambda) + (1 - \ell) (1 + \delta) H_{2}(\lambda),
\end{align}
where 
\begin{align}
H_{2}(\lambda) \equiv  \frac{1}{\sqrt{2\pi ( \alpha \bar{q} + \mu m^{2})} } \exp \left( - \frac{(\alpha/2 + \lambda)^{2}}{2 (\alpha \bar{q} + \mu m^{2})} \right) 
\end{align}
This is a linear equation for $\delta$ that can be solved to give:

\begin{align}
\delta = \frac{\ell H_{2}(-\lambda) + (1 - \ell) H_{2}(\lambda)}{1 - \ell H_{2}(-\lambda) - (1 - \ell) H_{2}(\lambda)}.\\
\end{align}

We also have in the zero temperature limit
\begin{align}
H_{1}(\lambda)  = \int \frac{1}{1 + e^{\Theta(z, \lambda)}} Dz = \int_{B(\lambda)/A_{0}} Dz &= \frac{1}{2} {\rm erfc} \left( \frac{B(\lambda)}{\sqrt{2} A_{0}}\right),\\
\end{align}

so that the zero temperature order parameter is

\begin{align}
\bar{q} &= \ell H_{1}(- \lambda) + (1 - \ell) H_{1}(\lambda),\\
\ell - \bar{\ell} & = \ell H_{1}(- \lambda) - (1 - \ell) H_{1}(\lambda).\\
\end{align}

Finally, we use the following small $T$ expansion:

\begin{align}
\int Dz \log \left(1 + e^{ - \Theta(z, \lambda)}\right) &= \frac{A_{0}}{\sqrt{2\pi}} e^{ - B(\lambda)^{2}/ 2 A_{0}} - B(\lambda) \frac{1}{2} {\rm erfc}( B(\lambda)/\sqrt{2} A_{0}),\\
& = \frac{1}{T(1 + \delta)}  \left[ (\alpha \bar{q} + \mu m^{2}) J_{2}(\lambda) - ( \frac{\alpha}{2} + \lambda) J_{1}(\lambda)\right] + O(1),
\end{align}

to write the zero temperature free energy density

\begin{align}
\lim_{T\to 0} f  =    \frac{\alpha \bar{q}   \delta }{2  (1 +\delta )^{2}}    +  \frac{ \mu m^{2}}{2} \frac{(1 + 2 \delta)}{ (1 + \delta)^{2}} + \frac{\lambda  (\ell - \bar{\ell})}{1+\delta} 
&- \frac{\ell}{(1 + \delta)}  \left[ (\alpha \bar{q} + \mu m^{2}) H_{2}(-\lambda) - ( \frac{\alpha}{2} - \lambda) H_{1}(-\lambda)\right]\\
&- \frac{(1-\ell)}{(1 + \delta)}  \left[ (\alpha \bar{q} + \mu m^{2}) H_{2}(\lambda) - ( \frac{\alpha}{2} + \lambda) H_{1}(\lambda)\right].
\end{align}

{\bf Summary of zero temperature MFT in lossy phase:} At zero temperature, we have

\begin{align}
\bar{q} &= \ell H_{1}(- \lambda) + (1 - \ell) H_{1}(\lambda) \label{eq:A},\\
\ell - \bar{\ell} & = \ell H_{1}(- \lambda) - (1 - \ell) H_{1}(\lambda) \label{eq:B}\\
\delta &= \frac{\ell H_{2}(-\lambda) + (1 - \ell) H_{2}(\lambda)}{1 - \ell H_{2}(-\lambda) - (1 - \ell) H_{2}(\lambda)}\label{eq:C}
\end{align}

with
\begin{align}
H_{1}(\lambda) &= \frac{1}{2} {\rm erfc} \left[ \frac{\alpha/2 + \lambda}{ \sqrt{2 ( \alpha \bar{q} + \mu m^{2})}} \right] , \quad H_{2}(\lambda) = \frac{1}{\sqrt{2 \pi ( \alpha \bar{q} + \mu m^{2})}} \exp \left( - \frac{(\alpha/2 + \lambda)^{2}}{2 (\alpha \bar{q} + \mu m^{2})}\right)
\end{align}

{\bf Solving zero temperature MFT}: Here we provide some details on how we find solutions for the MFT. We focus on the setting with $\mu = 0$, and in the lossy limit. Our goal is to simplify the MFT equations in this phase Eqs. \ref{eq:A}, \ref{eq:B},\ref{eq:C}. First, define 
\begin{align}
x \equiv H_{1}( - \lambda) = \frac{1}{2} {\rm erfc}\left( \frac{\alpha/2 - \lambda}{\sqrt{2 \alpha \bar{q}}}\right), \quad y \equiv H_{1}(\lambda).
\end{align}

From these, the constraint equation gives

\begin{align}
y(x) = \frac{1}{1 - \ell} \left( \ell x - \ell + \bar{\ell}\right) = \frac{\ell}{1 - \ell} \left( x - 1 + C\right),
\end{align}

and the order parameter becomes

\begin{align}
\bar{q} = \ell x + (1 - \ell) y(x) = \ell \left(2 x + C - 1\right). \label{eq:q_x_y}
\end{align}

Next, we use

\begin{align}
\frac{\alpha/2 - \lambda}{\sqrt{2 \alpha \bar{q}}} = {\rm erfc}^{-1}( 2 x), \quad  \frac{\alpha/2 + \lambda}{\sqrt{2 \alpha \bar{q}}} = {\rm erfc}^{-1}( 2 y(x)) . \label{eq:inverse}
\end{align}

It is important that we select the branch which has

\begin{align}
{\rm erfc}^{-1}( 2 x) + {\rm erfc}^{-1}( 2 y(x)) > 0.
\end{align}

which implies a constraint

\begin{align}
x < 1 - \bar{\ell}. 
\end{align}

In addition, we require $y(x) > 0$, which requires

\begin{align}
x > 1 - C.
\end{align}

from Eq. \ref{eq:inverse}, we get

\begin{align}
\bar{q} = \frac{\alpha}{2}  \left({\rm erfc}^{-1}( 2 x) +  {\rm erfc}^{-1}( 2 y(x))\right)^{-2},
\end{align}

combining this with Eq. \ref{eq:q_x_y} gives the implicit equation for $x \in [ 1- C, 1 - \bar{\ell}]$:

\begin{align}
(2 x + C - 1) = \frac{\alpha}{2 \ell}  \left({\rm erfc}^{-1}( 2 x) +  {\rm erfc}^{-1}( 2 y(x))\right)^{-2}. \label{eq:implicit}
 \end{align}

We see from this that solutions to $x$ depend on $\alpha$ only through the ratio $\alpha/\ell$. However, they do depend on the absolute value of $\ell$ as well, through the ratio $\ell/(1 - \ell)$.

Finally, an additional condition for a self-consistent solution is that $Q_{1} > 0$, This turns into the following inequality:

\begin{align}
1 - \ell \frac{1}{\sqrt{2 \pi \alpha \bar{q}}} e^{ - \left( {\rm erfc}^{-1} ( 2 x) \right)^{2}} -  (1-\ell) \frac{1}{\sqrt{2 \pi \alpha \bar{q}}} e^{ - \left( {\rm erfc}^{-1} ( 2 y(x)) \right)^{2}} > 0.
\end{align}
The critical curve is given by the implicit equation:

\begin{align}
\sqrt{2\pi \alpha (2 x + C - 1)/\ell} = e^{ - \left( {\rm erfc}^{-1} ( 2 x) \right)^{2}} +   \frac{(1-\ell)}{\ell} e^{ - \left( {\rm erfc}^{-1} ( 2 y(x)) \right)^{2}}.
\end{align}

We compute these curves numerically for different $\ell$ on the $\alpha$-$C$ plane in Fig. \ref{fig:max_alpha}, and plot them with solid colored lines. There is another transition for compression ratios closer to unity, indicated by dashed lines in Fig. \ref{fig:max_alpha}, in which there is a bifurcation in the solutions to \ref{eq:implicit}. This is therefore a topological transition: below the dashed curves, there is only one solution to \ref{eq:implicit}, whereas above it there are three. The point where the solid curve hits the dashed curve (shown with a star in the figure) represents the maximal value of $\alpha$ for which lossless compression can occur. In the right panel of Fig. \ref{fig:max_alpha}, we show that this maximum value depends on the total message length $\ell$, but has an upper bound and never exceeds $\alpha \approx 0.4049 $ for all $\ell$. We argue that this value has to occur at $\lambda = 0$. In this case, we get ${\rm erfc}^{-1} y(x) = {\rm erfc}^{-1} x$, $\bar{q} = x$,  $C = 1 - x$, and we must solve both of the following equations simultaneously to find this maximal $\alpha$:

\begin{align}
\sqrt{2 \pi \alpha x} =  e^{ -  \left( {\rm erfc}^{-1} ( 2 x) \right)^{2}} , \quad x  = \frac{\alpha }{4 \left( {\rm erfc}^{-1}(2x)\right)^{2}}.
\end{align}

\subsection{Zero Compression Limit: $C = 1$}

\paragraph{Half Filling}

The easiest setting to consider is half-filling, i.e. $\ell = 1/2$. In this case, with $C = 1$, the constraint equation implies that $\lambda = 0$, and that

\begin{align}
\bar{q} = \frac{1}{2} {\rm erfc} \left( \frac{\alpha}{2 \sqrt{2 \alpha \bar{q}}}\right)
\end{align}

This has nonzero solutions up to $\alpha \approx  0.6629$. Above this, the only solution is $\bar{q} = 0$, which also has zero entropy. This corresponds to zero distortion. In the compressed sensing problem, this is akin to perfect reconstruction. Below this limit, the finite entropy solution for $\bar{q}$ has nonzero distortion.

We may also explore the finite $\Delta$ solutions in this limit. We get the system of equations

\begin{align}
0 & =  \ell \int Dz \frac{1}{1 + e^{\Theta_{-}(z)}} - (1 - \ell) \int Dz \frac{1}{1 + e^{ \Theta_{+}(z)}} , \\
q_{0} &= \ell \int Dz \frac{1}{1 + e^{\Theta_{-}(z)}} + (1 - \ell) \int Dz \frac{1}{1 + e^{ \Theta_{+}(z)}}, \\
q &= \ell \int Dz \frac{1}{\left(1 + e^{\Theta_{-}(z)} \right)^{2}} + (1 - \ell) \int Dz \frac{1}{\left(1 + e^{ \Theta_{+}(z)}\right)^{2}}.
\end{align}

where
\begin{align}
\Theta_{\pm}(z) = - \frac{1}{ \Delta} \sqrt{ \alpha q } z + \frac{\alpha }{2 (1 +  \Delta)} \pm X,
\end{align}

and the free energy density at zero temperature is $f = 0$.

\paragraph{Arbitrary Filling}

Now consider arbitrary $\ell$. This requires keeping $\lambda$. However, some simplifications occur. For instance, 

\begin{align}
\ell H_{1}(- \lambda) = (1 - \ell) H_{1}(\lambda)
\end{align}
which implies

\begin{align}
\bar{q} = 2 \ell H_{1}(-\lambda)
\end{align}

In general, the transition to perfect reconstruction (above which the only solution is $\bar{q} = 0$) occurs at

\begin{align}
\alpha/\ell \approx 1.3257
\end{align}

For larger values of $\alpha$, $\bar{q} = 0$. This means that for $P/ L > 1.35$, the typical configuration recovers the original message exactly. This occurs when the dimensionality of the embedding space is larger than the message length, suggesting that it is a limit in which the message vectors are statistically orthogonal. In the opposite limit, $P< L$, the message must utilize many linearly dependent vectors to construct the message.

\begin{figure*}[htbp!]
\includegraphics[width = \textwidth]{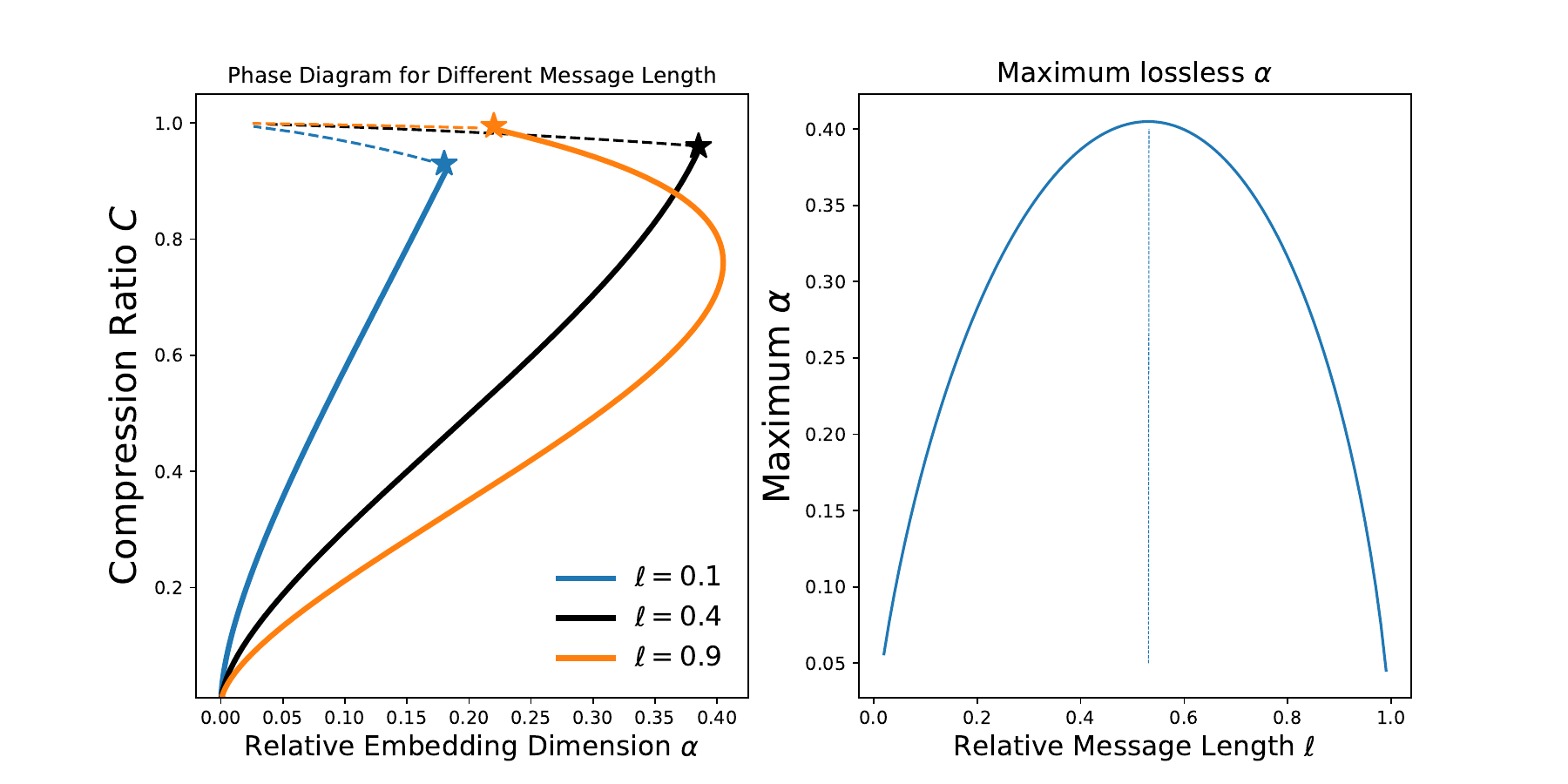}
	\caption{ {\it Phase diagram dependence on message length} {\bf (Left)}  The zero temperature phase diagram of the RS MFT for different values of relative message length $\ell$. In each figure, we plot only the curves demarcating the compressible phase $\Delta>0$. The solid curve indicates the discontinuous transition, whereas the dashed curve corresponds the appearance of self-consistent incompressible ($\Delta = 0$) solutions. For each fixed $\ell$, the compressible region extends out to some maximal $\alpha$ corresponding to the point where these two curves meet (denoted by a star in the figure). {\bf (Right)}  shows that the maximal $\alpha$ depends non-monotonically on $\ell$, and tends to zero both as $\ell \to 0$ and $\ell \to 1$.    }
	\label{fig:max_alpha}
\end{figure*}

\section{Annealed Calculation}\label{app:annealed}

Here we calculate the annealed free energy

\begin{align}
f_{\beta}^{a} = - \frac{1}{\beta N} \log \mathbb{E}\left[ Z_{\beta}(c, E)\right]_{c, E},
\end{align}
from which we derive the expression for the lower bound on the minimal distortion. We specialize to centered Gaussian embeddings with unit variance. In this case, the disorder averaged partition function can be directly evaluated

\begin{align}
\bar{Z}_{\beta} \equiv  \mathbb{E}\left[ Z_{\beta}(c, E)\right]_{c, E} = \sum_{R = 0}^{\bar{L}} \Omega(R) \left(1 + \beta q(R)\right)^{- P/2},
\end{align}

where $q(R) = \ell + \bar{\ell} - 2 R/N$, and 

\begin{align}
\Omega(R) = \left( { L \atop R}\right) \left( { N - L \atop \bar{L} - R}\right)
\end{align}
gives the total number of configurations which satisfy $\sum_{i = 1}^{N} c_{i} \bar{c}_{i} = R$. We now evaluate this partition function using Laplace's method. In the scaling limit considered in this paper, where $r = R/N$, 

\begin{align}
\Omega(r) \approx e^{N S(r)}, \quad S(r) = \ell h\left(r/\ell \right) + (1 - \ell) h \left( \frac{\bar{\ell} - r}{1 - \ell}\right), \quad h(p) = - p \log p - (1 - p) \log (1 - p),
\end{align}
in which case we get

\begin{align}
\bar{Z}_{\beta} \approx N \int_{0}^{\bar{\ell}} dr e^{ N \Phi(r,\beta)}, \quad \Phi(r, \beta) = S(r) - \frac{\alpha}{2} \log \left( 1+  \beta ( \ell + \bar{\ell} - 2 r) \right).
\end{align}

For large $N$, we'll need the first two derivatives of the function $\Phi(r)$. Let $\beta = \tau^{-1}$, then

\begin{align}
\Phi_{r} & = \frac{\alpha}{\tau + \ell + \bar{\ell} - 2r}  + \log \left( \frac{(\ell - r)(\bar{\ell} - r)}{r (1 - \ell - \bar{\ell} + r)}\right),\\
\Phi_{\tau} & =  \frac{\alpha}{2} \frac{1}{1 + q/\tau} \frac{q}{\tau^{2}} = \frac{\alpha}{2 \tau } \frac{q}{\tau +  q},\\
\Phi_{rr} &= \frac{2 \alpha}{( \tau + \ell + \bar{\ell} - 2 r)^{2}} + \frac{1}{r - \bar{\ell}} + \frac{1}{r ( r - \ell)}  - \frac{1}{1 - \ell - \bar{\ell} + r}.
\end{align}

Then

\begin{align}
\bar{Z}_{\tau} \approx \sum_{i} e^{ N \Phi(r_{i}^{*})} \int d \delta r e^{ \frac{1}{2} N \Phi_{rr}(r_{i}^{*}) \delta r^{2}}  .
\end{align}

The annealed free energy will be determined by the dominant stable saddle

\begin{align}
f^{a}_{\tau} = - \tau \Phi(r^{*}(\tau), \tau), \quad \Phi_{r}(r^{*}(\tau), \tau) = 0, \quad \Phi_{rr}(r^{*}, \tau) < 0. \label{eq:annealed_f}
\end{align}

The bound on the minimum distortion follows by maximizing the annealed free energy over $\beta$. The logic follows in a few lines:

\begin{align}
 D_{min} = \mathbb{E}\left[ \frac{ 1}{  N} \min_{\bar{c}} H(\bar{c}, c)  \right]_{c,E} \geq \mathbb{E}\left[-\frac{1}{\beta N} \log Z_{\beta}\right] \geq -\log \mathbb{E}\left[ Z_{\beta}\right] .
\end{align}
The first inequality comes by noting that $e^{ - \beta H(\bar{c}, c)} \le e^{ - \beta H_{min}}$ for all $\bar{c}$, and the second inequality comes from Jensen's inequality.  Since this inequality is true for all $\beta$, we may as well make the lower bound as good as possible and maximize the RHS over $\beta$. Therefore, we get the distortion density lower bound:

\begin{align}
D_{min}  \ge \max_{\beta} f^{a}_{\beta}.
\end{align}

At fixed $\alpha, \ell, \bar{\ell}$, we find the temperature $\tau$ which maximizes the free energy. This will satisfy:

\begin{align}
\partial_{\tau} f = - \Phi(r^{*}(\tau), \tau) - \tau \Phi_{\tau}(r^{*}(\tau), \tau) = 0,
\end{align}
which implies
\begin{align}
\Phi = - \tau \Phi_{\tau} = -\frac{\alpha q \beta}{2 (1 + \beta q)}.
\end{align}
Plugging this into the expression for the annealed free energy \ref{eq:annealed_f} gives the form presented in the main text. 
\begin{align}
f_{\beta^{*}}(r^{*}) =  \frac{ \alpha q}{2 (1 + \beta^{*} q)} .
\end{align}

\end{widetext}

\end{document}